# Controlling local resistance via electric-field induced dislocations

D. M. Evans,[1] D. R. Småbråten,[1] T. S. Holstad,[1] P. E. Vullum,[2] A. B. Mosberg,[3] Z. Yan,[4,5] E. Bourret,[5] A. T. J. Van Helvoort,[3] S. M. Selbach,[1] and D. Meier[1]

[1]Department of Materials Science and Engineering, Norwegian University of Science and Technology (NTNU), Trondheim, Norway
[2]SINTEF Industry, 7491 Trondheim, Norway
[3]Department of Physics, Norwegian University of Science and Technology (NTNU), Trondheim, Norway
[4]Department of Physics, ETH Zurich, 8093 Zürich, Switzerland
[5]Materials Sciences Division, Lawrence Berkeley National Laboratory, Berkeley, California 94720, USA

E-mail: donald.evans@ntnu.no and dennis.meier@ntnu.no

**Dislocations are one-dimensional (1D) topological line defects where the lattice deviates from the perfect crystal structure. The presence of dislocations transcends condensed matter research and gives rise to a diverse range of emergent phenomena[1–6], ranging from geological effects[7] to light emission from diodes[8]. Despite their ubiquity, to date, the controlled formation of dislocations is usually achieved via strain fields, applied either during growth[9,10] or retrospectively via deformation, e.g., (nano[11–14])-indentation[15]. Here we show how partial dislocations can be induced using local electric fields, altering the structure and electronic response of the material where the field is applied. By combining high-resolution imaging techniques and density functional theory calculations, we directly image these dislocations in the ferroelectric hexagonal manganite Er(Ti,Mn)O$_3$ and study their impact on the local electric transport behaviour. The use of an electric field to induce partial dislocations is a conceptually new approach to the burgeoning field of emergent defect-driven phenomena and enables local property control without the need of external macroscopic strain fields. This control is an important step towards integrating and functionalising dislocations in practical devices for future oxide electronics.**

Classically, dislocations have often been regarded as an imperfection in otherwise perfectly ordered crystals that are detrimental to electronic functionality[16,17]. However, very recently there has been growing interest in the properties of individual dislocations *because* they have different symmetries, and therefore properties, to the surrounding material[18]. Examples of emergent functional physical properties include locally enhanced conductivity[19], ferromagnetic order in antiferromagnets[20], redox-based resistive switching behaviour[4], and trapping of light[21]. Due to the low dimensionality of dislocations – which makes them highly stackable – and their unique properties, dislocations are now perceived as promising atomic-scale entities for next-generation device applications[17,18]. Despite their outstanding application potential, there are key limitations to the technological development of dislocations: Most critical is the control of their formation. Currently, dislocations are created by strain engineering, predominantly in one of two ways: Either strain fields during growth, so that dislocations form to release the strain[9,10,22], or post-growth via applied stress[15], e.g., by nanoindentation[11–14]. Although strain engineering is a highly efficient tool for creating dislocations, it is challenging to alter the properties of a material exactly via strain fields at the local scale; particularly problematic after a material has been implemented into a device architecture.

Here, we use electric fields to create partial dislocations in a ferroelectric material with nanoscale spatial precision, altering the structure and electronic transport behaviour where the field was applied. The hexagonal manganite $Er(Mn,Ti)O_3$ is chosen as model system (see Methods for details)[23,24], but similar behaviour is expected in structurally equivalent systems such as hexagonal gallates, indates, and ferrites. The ferroelectric domain structure of (110)-oriented $Er(Mn,Ti)O_3$ is shown in the piezoresponse force microscopy (PFM) image in Fig. 1**a**. The PFM data demonstrates that the two possible polarisation domains, $+P$ (grey) and $-P$ (dark), form six-fold meeting points that are characteristic for hexagonal manganites[25,26]. After mapping the domain structure of $Er(Mn,Ti)O_3$, we apply $-60$ V (for 5 s) to the back electrode while keeping the probe tip static. Using a simple point charge approximation ($E = V/r_{tip}$), the biased static tip generates an electric field of about 6 MV/cm [refs. [27,28]]. Subsequent conductive atomic force microscopy (cAFM) images (taken with +45 V) reveal a local modification of the transport behaviour, manifesting as an ellipsoidal area of enhanced conductance (gold, in Fig. 1**b**) with an insulating core: The core of reduced conductance occurs directly below the position of the AFM tip, i.e., the position of the maximum electric field. A scanning electron microscopy (SEM) image of the same region as in Fig. 1**b** is presented in Fig. 1**c**, which demonstrates that the area of enhanced conductance in the cAFM scan correlates with bright contrast in SEM, associated with an increased yield of secondary electrons. A direct comparison of the SEM and cAFM images is given by the cross-sectional profiles in Fig. 1**d**, showing excellent correlation. Figure 1 thus reveals that the electrically modified region displays two regimes: (i) a broader area with enhanced conductance due anti-Frenkel defects as reported in ref. [[29]], and (ii) a highly localized area with reduced conductance. In contrast to the enhanced conductance in regime (i), the suppressed conductance in regime (ii) cannot be explained by the same defect formation process[29], indicating a different microscopic origin.

To investigate the underlying microscopic effect, we study the atomic structure in the electrically modified region, using transmission electron microscopy (TEM) techniques on a focused ion beam (FIB) prepared lamella (see Methods). Figure 2**a** shows a high-angle annular-dark-field scanning transmission electron microscopy (HAADF-STEM) image of a cross-section centred at the electrically modified region presented in Fig. 1. The bright dots correspond to Er and Mn atomic columns; the heavier Er atoms give brighter dots than the lighter Mn atoms. The HAADF-STEM shows that the Er layers are disrupted in several places and appear to be displaced, seemingly merging with the Mn layers in the projected image. To visualise associated local changes in the lattice periodicity, Fourier filtering is applied as presented in Fig. 2**b**. This treatment reveals discontinuities in the lattice periodicity and the presence of edge dislocations, marked by the green and red circles in Fig. 2**b**. An enlarged section with higher resolution of the feature in the red box in Fig. 2**a** is shown in Fig. 2**c**. On the left side of the dislocation in Fig. 2**c**, the HAADF-STEM image shows the normal sequence of alternating Er and Mn layers, whereas the right side of the image shows Er columns that continuously merge into Mn columns and vice versa. We note there is no clear interface where Er columns swap with Mn columns, as the interface is inclined to the [001] viewing direction. The general possibility to stabilize such stacking faults is not surprising[30] as their growth-induced formation has been reported in hexagonal manganites and ferrites[31–33]. In contrast to previous work, however, the stacking faults in our system are generated post-growth and with nanoscale spatial precision by the application of an electric field.

To evaluate the stability of the defect structure resolved in Fig. 2**c**, we model the dislocated lattice using density functional theory (DFT) as illustrated in Fig. 2**d**. To look for stable states, we consider the isostructural system YMnO$_3$ [refs. [34,35]], where multiple 240 atom 1x8x1 supercells were initialized in the high-symmetry $P6_3/mmc$ phase with different dislocation configurations and relaxed into the polar state $P6_3cm$, until the forces on all the atoms were below 0.01 eV Å$^{-1}$. The DFT calculations show that a stable dislocated structure exists when the lattice is displaced by *c*/4 and *a*/3 (or one of the 6 symmetry equivalent directions, perpendicular to the unique axis), as shown in Fig. 2**d**. The calculated dislocation configuration is in excellent agreement with our HAADF-STEM data (Figure S1 shows the dislocated structure viewed down the [110] direction, which is symmetry equivalent to [$\bar{1}$00]), leading us to the conclusion that the merging features in Fig. 2**c** correspond to a stacking fault between two partial dislocations.

Based on the DFT calculations, we can predict the expected atomic pattern when we image along other crystallographic directions, allowing us to conduct an independent test experiment. The calculated crystal structure viewed along different directions, with and without the dislocated structure, is presented in Fig. S2, including the superposition of the dislocated structure and the unperturbed Er(Ti,Mn)O$_3$ structure with space group symmetry *P6$_3$cm* [36]. This superposition accounts for the depth convolution present within TEM lamellas with finite thickness. A representative HAADF-STEM image recorded along the [001] direction of an Er(Ti,Mn)O$_3$ lamella with electrically modified regions (analogous to Fig. 1) is presented in Fig. 3**a**. The HAADF-STEM image shows the Er atoms as bright dots, reflecting the characteristic pattern associated with the *P6$_3$cm* space group symmetry of the crystal sketched in Fig. S2. The latter is confirmed by the high-resolution HAADF-STEM image in Fig. 3**b**, taken from the area marked red in Fig. 3**a**. On a closer inspection, however, local deviations from the ideal structure are observed, as shown by the HAADF-STEM image in Fig. 3**c** (obtained in the area marked green in Fig. 3**a**). Here, the Er atoms are found to form a close-packed honeycomb-like structure as illustrated by schematic overlay, in accordance with the DFT-predicted structure (see Fig. S2; blue and gold dots in the overlay represent Er atoms from volumes on either side of the stacking fault).

After establishing the presence of the structural defects in the region of reduced conductance, we calculate the associated formation energy. For the partially dislocated structure in Fig. 2**d**, we find a formation energy of about 755 mJ m$^{-2}$, which is only ~7 times higher than for the charged domain walls, which naturally occur in the hexagonal manganites[37]. The DFT calculations further show that the dislocated structure significantly alters the electrostatic potential energy. To quantify this, we calculate the planar average across the supercell in Fig. 4**a**, which shows a decrease in the potential energy of $\approx$ 0.75 eV ($\approx$ 1/2 E$_g$) compared to the bulk, corresponding to bound negative charges associated with the stacking fault. However, in contrast to the negatively charged tail-to-tail domain walls in hexagonal manganites[37], neither calculated Bader charges, nor Mn magnetic moments, reveal any inherent electronic charge transfer between the dislocated structure and the bulk. This is because the magnitude of the band bending (visualised in Fig. S3) does not fulfil the Zener-like breakdown criterion[37], meaning that our defects are not compensated by electronic charge carriers.

However, the bound negative charges associated with the dislocated structure in Fig. 4**a** can attract mobile ionic defects with a relative positive charge, such as oxygen vacancies ($v_O^{\bullet\bullet}$). In fact, we find that the $v_O^{\bullet\bullet}$ formation energy is reduced by about 0.15 eV compared to the bulk (Fig 4**b**), demonstrating a significant defect segregation enthalpy. To contextualise the strength of this driving force: it is five times higher than the attraction of oxygen interstitials ($O_i^{''}$) to neutral domain walls in the hexagonal manganites ($\approx 0.03$ eV), which are experimentally well documented to have enhanced conductance[38]. As such, the five times higher driving force is expected to pin $v_O^{\bullet\bullet}$ to the stacking faults and vice versa. To illustrate the corresponding changes in electronic state, we calculate the density of states (DOS) in YMnO$_3$ with and without $v_O^{\bullet\bullet}$ at the dislocated structure, Fig. 4**c**. The DOSes are qualitatively similar, but the $v_O^{\bullet\bullet}$ DOS has a strongly localized state within the bandgap, and a higher Fermi energy. This increase in Fermi energy at the interface gives a variation in electronic charge carrier concentration and – in p-type system with holes as majority carriers[35] – is expected to cause a diffusion current, leading to a local reduction in the number of available electronic charge carriers. As a consequence, the dislocation-induced structural changes enhance the resistivity relative to the bulk, which is consistent with the lower conductance observed in the cAFM data in Fig. 1**b**.

Our work demonstrates that electric fields can be used to create partial dislocations in a complex oxide to engineer the local structure and electronic response, bypassing the necessity of applied strain fields. The ability to induce such changes post-growth, on demand, and with nanometre spatial precision provides a conceptually new approach to local property engineering as exemplified by the enhanced resistance in hexagonal manganites. Our approach is expected to be generally applicable to dielectric materials close to a structural instability, where a sufficiently large electric field can build-up. Because the electric-field driven dislocation injection is applicable after a material has been synthesized or integrated into a device, it presents a new opportunity for engineering functional materials via low-dimensional structural defects.

**Acknowledgements:** The authors wish to thank Knut Marthinsen and Jan Schultheiß for helpful and insightful discussions about dislocations. DRS and SMS were supported by the Research Council of Norway (FRINATEK project number 231430/F20 and 275139/F20) and acknowledge UNINETT Sigma2 (Project no. NN9264K) for providing the computational resources. ABM was supported by NTNU's Enabling technologies: Nanotechnology. The Research Council of Norway is acknowledged for the support to the Norwegian Micro- and Nano-Fabrication Facility, NorFab, project number 245963/F50 and Norwegian Centre for Transmission Electron Microscopy, NORTEM, Grant No. 197405. ZY and EB were supported by the U.S. Department of Energy, Office of Science, Basic Energy Sciences, Materials Sciences and Engineering Division under Contract No. DE-AC02-05-CH11231 within the Quantum Materials program-KC2202. DM thanks NTNU for support through the Onsager Fellowship Programme and NTNU Stjerneprogrammet.

**Author contributions** DME coordinated the project and lead the scanning probe microscopy work together with TSH, both supervised by DM. ABM was supervised by AvH and conducted the FIB and SEM work. PEV and ABM conducted the TEM work. DRS performed the DFT calculations supervised by SMS. ZY and EB provided the materials. DME and DRS analysed and interpreted the data, under the supervision of DM and SMS, respectively. DME and DM wrote the manuscript. All authors discussed the results and contributed to the final version of the manuscript.

# Figures and Figure Captions

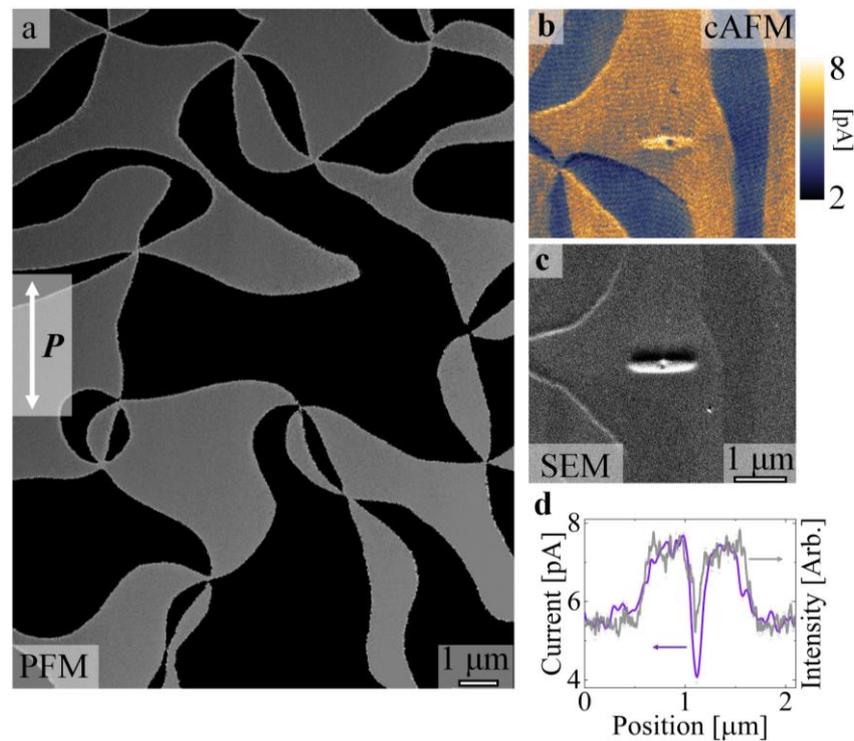

**Figure 1 | Electric-field control of local conductance in Er(Ti,Mn)O$_3$. a,** Lateral DART mode PFM phase image of an in-plane ferroelectric domain pattern, taken with 5 V and a frequency range of 784.5-794.5 kHz. Bright and dark contrasts represent ferroelectric 180° domains, revealing the characteristic domain pattern of hexagonal manganites. **b**, cAFM scan recorded with +45 V on the same sample after the application of -60 V for 5 s, to the back electrode with a stationary AFM tip; the position where the top was placed coincides with the blue centre of the gold elliptical feature seen in the cAFM map. Bright (gold) colours correspond to areas of higher conductance, while dark (blue) colours correspond to areas of lower conductance. **c**, SEM image of the feature in **b** demonstrating conductivity-sensitive SEM contrast. **d,** Cross sectional graph, correlating SEM contrast and cAFM contrast.

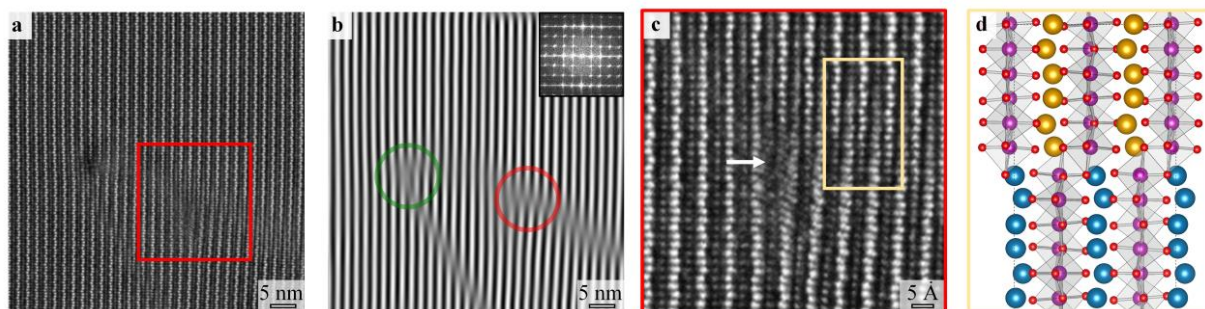

**Figure 2 | Atomic-scale structure of electrically induced dislocations and stacking faults.
a,** HAADF-STEM image viewed down the [$\bar{1}$00] direction, showing the region beneath the position of the AFM tip during application of the electric field. The periodic crystal structure is interrupted by features extending in directions close to the <011> directions. The bright and grey dots are Er and Mn atomic columns, respectively. **b**, Inverse fast Fourier transformation (FFT) obtained by selecting only the (002) maxima of an FFT of **a**, see insert. The image shows areas with the same periodicity of the Er lattice, allowing easier identification of lattice defects. Two edge dislocations are highlighted by red and green circles. **c,** Representative HAADF-STEM image taken across the crystallographic features, corresponding to the area marked by the red box in **a**. **d,** Fully relaxed DFT supercell, modelling the dislocated structure by unit cells dislocated by *c*/4 and *a*/3 with respect to each other. Large gold and blue spheres represent Er atoms on either side of the stacking fault; Mn and O atoms are sketched in purple and red, respectively.

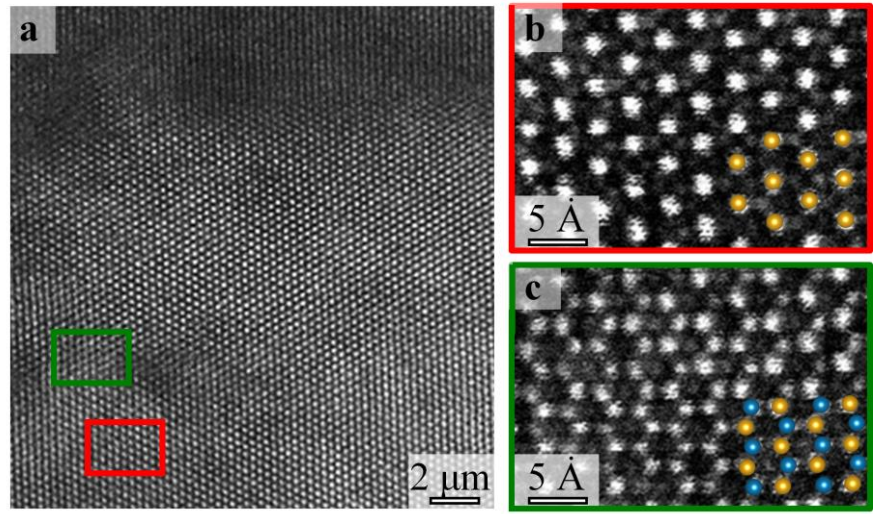

**Figure 3 | Atomic defect structure viewed along the [001] axis. a,** HAADF-STEM image a region with electric-field induced defects, written with the same parameters as used in Fig. 1 and Fig. 2. Bright dots correspond to Er atom columns. **b,** HAADF-STEM image of the red box area of **a**, showing the expected *R*-cation lattice for a hexagonal crystal when viewed along the [001] direction. The gold dots correspond to the *R*-cations from a DFT simulation of the structure, looking down the [001] axis. Mn columns can be seen as weak dots in between the more pronounced dots from the Er columns. **c,** HAADF-STEM image corresponding to the area marked by the green box in **a**, showing Er pattern that deviates from the unperturbed crystallographic structure seen in **b**. Gold and blue dots represent the bulk and shifted Er atom columns when the stable dislocation of Figure 2 is viewed down the [001]. The overlay shows that the dislocated structure reproduces the pattern resolved by HAADF-STEM.

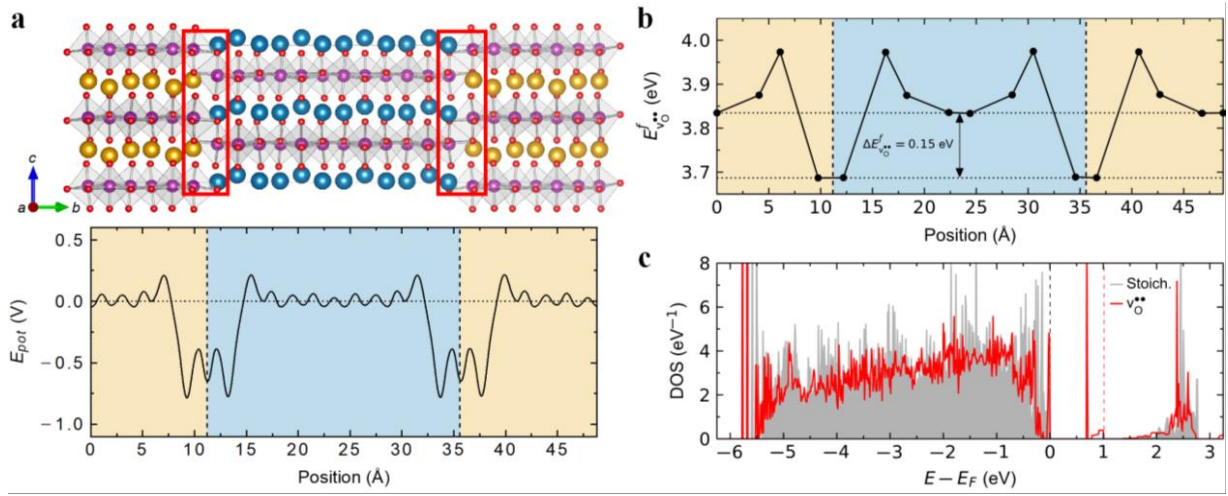

**Figure 4 | Electronic defect structure calculated. a,** Dislocated structure (upper part) and electrostatic potential (lower part) calculated by density functional theory (DFT). The calculated electrostatic potential is scaled relative to the potential away from the dislocated interfaces, which are indicated by the red boxes. The significant reduction in the potential (ca.1/2 $E_g$ at the interface) corresponds to bound negative charges, which are likely to attract positively charged $v_O^{\bullet\bullet}$. **b,** $v_O^{\bullet\bullet}$ formation energy with increasing distance from the interface(s), showing a significant energy preference, 0.15 eV, for forming $v_O^{\bullet\bullet}$ at the interfaces(s) compared to the bulk. **c,** Calculated local DOS at the partial-dislocations in a stoichiometric supercell (grey), and with an accumulation of oxygen vacancies (red). The corresponding Fermi levels for the pristine cell and the $v_O^{\bullet\bullet}$ defect cell are marked by dashed grey and red lines in **c**, respectively. The DOS are aligned to core-states.

**Methods**

Samples: Single crystals of $Er(Mn_{1-x},Ti_x)O_{3+\delta}$ with x = 0.002 were grown using the pressurized floating zone method [ref. 24]. From this, we prepared samples with the ferroelectric polarisation parallel to the sample surface (in-plane *P*). The crystals were orientated by Laue diffraction, cut to have thicknesses of ≈ 0.5 mm and electro-mechanically polished to give an RMS surface roughness of about 1 nm.

Focused Ion Beam (FIB): TEM specimen preparation was carried out using a Thermo Fisher Scientific Helios G4 UX Dual Beam Focused Ion Beam. Lamellas were prepared 'flipped' with *in-situ* lift-out and backside milling. Final polishing was carried out at 2 kV. High-resolution Pt markers were used in conjunction with C protection layers to ensure the final lamellas were centred on the non-conductive region.

Transmission electron microscopy (TEM) analysis was performed using a double Cs aberration corrected cold FEG JEOL ARM 200FC, operated at 200 kV. STEM imaging was performed with a 27 mrad semi convergence angle.

Atomic Force Microscopy: The lateral DART-PFM was performed on a Cypher ES Environmental AFM using an Oxford Instruments Asylec-01-R2 Pt/Ir tip. The remaining SPM measurements were performed on a NT-MDT NTEGRA SPM using a TipsNano DCP20 tip. The voltage was applied to the sample back-electrode with the tip connected to ground.

Density functional theory (DFT): Calculations on isostructural $YMnO_3$ (isostructural and electronically similar to $ErMnO_3$, but without complicating f-electrons) were carried out using the projector augmented wave (PAW) method as implemented in VASP,[39–41] using the Y_sv Mn_sv and standard O pseudopotentials supplied with VASP. PBEsol+U[42,43] with U = 5 eV on Mn 3d states, combined with a collinear frustrated antiferromagnetic order on the Mn sublattice, were used to reproduce the experimental lattice parameters[44] and electronic band gap.[45] The energy cut-off of the plane-waves was set to 550 eV. The Brillouin zone was sampled with a Γ-centered 4x1x2 k-point grid for geometry optimization and 6x1x3 for density of states (DOS) and electrostatic potential calculations. 240 atom 1x8x1 supercells were initialized in the high-symmetry $P6_3/mmc$ phase with different dislocation configurations and relaxed until the forces on all the atoms were below 0.01 eV Å$^{-1}$. Ferroelectric polarizations were calculated from a simple point charge model using formal charges[46]. The dislocation formation energy was calculated as

$$E^f = \frac{1}{2A}\left(E^f_{dis} - E^f_{ref}\right)$$

Where $E^f_{dis}$ and $E^f_{ref}$ are the energies of a dislocation supercell and a monodomain supercell, respectively, and $A$ is the cross-sectional area.

**Supplementary Information**

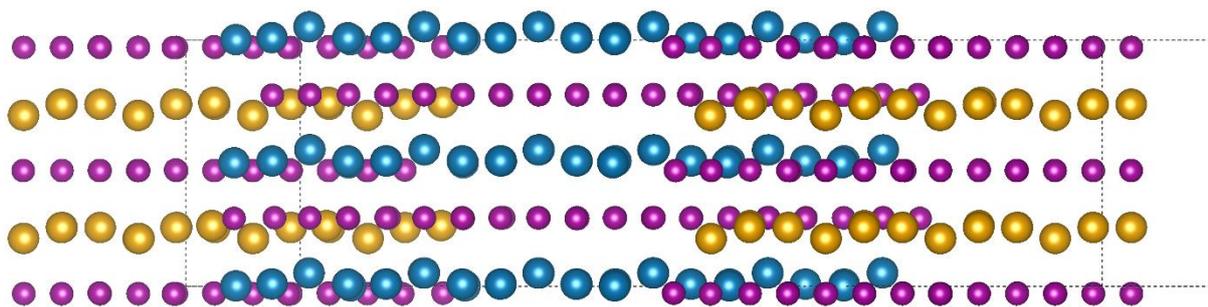

**Figure S1 | A *c*/4 and *a*/3 dislocated structure viewed at 60 º to the interface.** DFT calculations showing one of the symmetrically equivalent *c*/4, *a*/3, dislocated structures viewed along the [110] direction, or symmetrically equivalent projections. In this case, there is a continual merging of the *R*-cation and Mn atomic columns, as observed in the HAADF-STEM of Figure 2**c**.

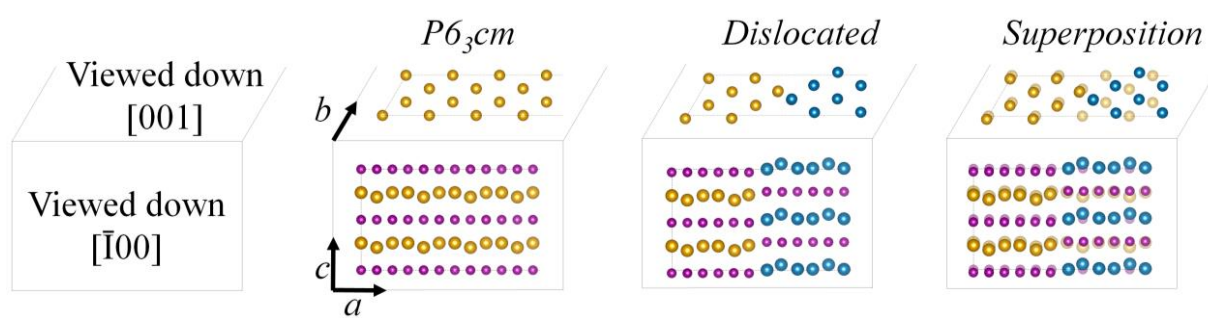

**Figure S2 | Viewing directions and depth effects.** Left: Illustration of the atomic structure in the unit cell of hexagonal manganites with space group symmetry *P6₃cm*, viewed along the [001] and [1̄00] directions. Er and Mn atoms are represented by gold and purple spheres, respectively. Middle: Comparison of dislocated structures; Er (gold and blue) and Mn (purple). Right: Superposition of the structures seen on the left and in the middle.

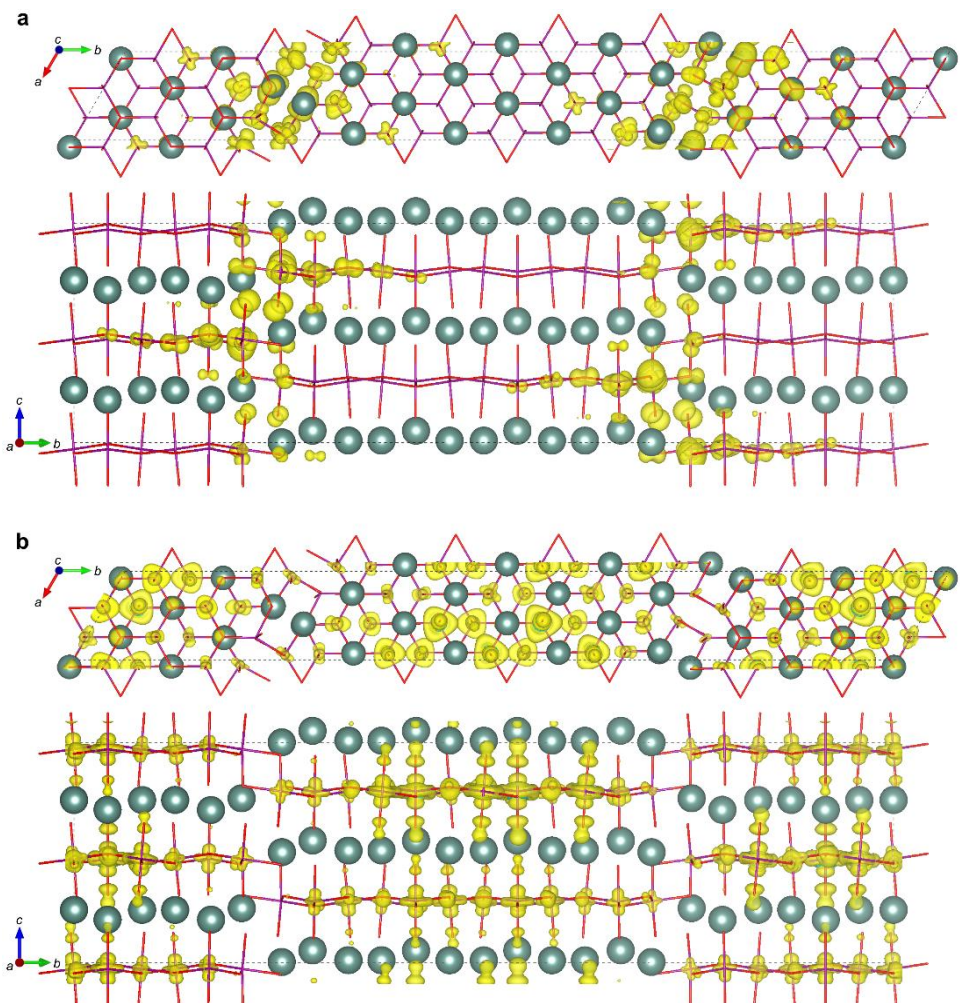

**Figure S3 | Change in local orbitals at a dislocation. a**, Visualization of the calculated partial charge density for the highest occupied wavefunction, illustrating the position of the valence band edge in the supercell. **b**, Visualization of the calculated partial charge density of the lowest unoccupied wavefunction, illustrating the position of the conduction band edge in the supercell.